\documentstyle[11pt,newpasp,twoside,epsf]{article}
\def\edcomment#1{\iffalse\marginpar{\raggedright\sl#1\/}\else\relax\fi}
\reversemarginpar

\begin{document}
\title{History of Star Formation in the Universe}
 \author{Bruce G. Elmegreen}
\affil{IBM Research Division, T.J. Watson
Research Center, PO Box 218, Yorktown Hts., NY 10598 USA}

\begin{abstract}
The cosmic history of star formation is briefly reviewed, starting
with the Milky Way and then discussing observations relevant to the
closed box and hierarchical build-up models.  Observations of local
star formation are reviewed for comparison.  The halo globular clusters
appear to have formed in about the same way as today's clusters, with
the same IMF and stellar densities, and with a larger mass reflecting
the higher pressures of their environments.  The bulge of the Milky Way
appears to have evolved according to the dissipational collapse model,
but other bulges may differ. 
\end{abstract}

to be published in "Chemical enrichment of the
intracluster and intergalactic medium,"
eds. F. Matteucci \& R. Fusco-Femiano, ASP Conference Series, 
in press 2001 (a conference at Vulcano Italy, 14-18 May 2001).

\section{Overview}

Surveys of the Hubble Deep Field and other regions have demonstrated over
the last few years that galaxies at high redshift are mostly spheroidal
until approximately $z\sim2$, and then disks form more recently (Madau
et al. 1996; Steidel et al. 1996; Giavalisco, Steidel \& Macchetto 1996;
Cowie et al. 1997; Franceschini et al. 1998; Rodighiero et al. 2000).
Theory suggests that dark matter halos also formed early, perhaps at
$z\ge3$ (Percival, Miller \& Peacock 2000).

The star formation rate during the spheroid epoch is about constant and
higher than during the average disk epoch by a factor of $\sim3$. The
star formation rate today is smaller still, perhaps by another factor
of 3. Figure 1 shows the cosmic history of star formation according
to  Boselli et al. (2001), compiled from various sources.  Cosmological
models explain the general trends in this diagram by having more galaxy
interactions and triggered starbursts at high z, and more isolated
disk star formation at low z (e.g., Rowan-Robinson et al. 1997; Baugh
et al. 1998; Tan, Silk, \& Ballard 1999; Cole et al. 2000; Nagamine,
Cen \& Ostriker 2000).

\begin{figure}
\plotone{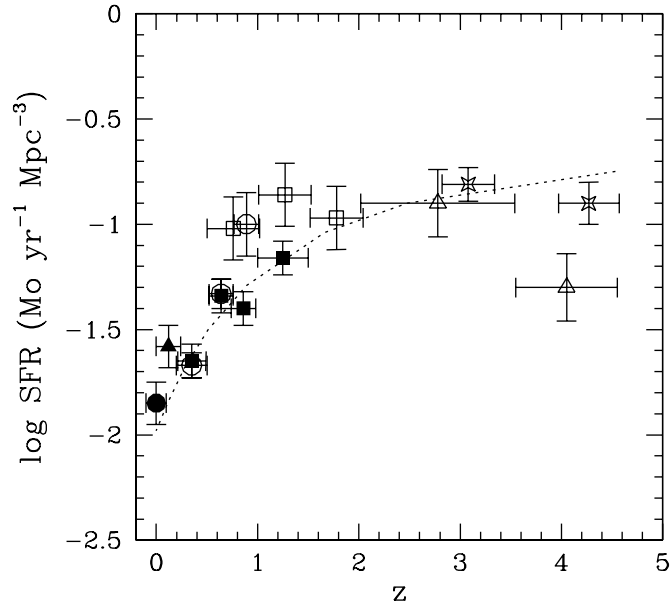}
\caption{Cosmic history of star formation from Boselli et al. (2001),
taking data from various sources and fitting a closed-box model to it.}
\end{figure}

\section{Milky Way Formation}

The sequence of events during the formation of the Milky Way gives a
consistent picture. The ages of the oldest halo globular clusters are
$12.9\pm2.9$ Gy (Carretta et al. 2000), corresponding to $z\sim3-4$ in
standard $\Lambda$CDM cosmologies.  The age of the solar neighborhood
is $\sim11$ Gy (Binney et al. 2000), which corresponds to $z\sim2$,
and the bulge formed between these two times, at $z\sim2-3$.

Most of what we know about the earliest star formation comes from studies
of globular clusters in our Galaxy and other galaxies. We have learned,
for example, that the oldest globular clusters probably formed before the
galaxies condensed. These blue globular clusters, which have the lowest
metallicities, have spatial distributions and velocity dispersions that
correlate better with the group potentials around elliptical galaxies
than with the potentials of the galaxies themselves (e.g., Blakeslee
et al. 1997; Minniti et al. 1998; Harris, Harris, \& McLaughlin 1998;
Kissler-Patig et al. 1999; see review in Elmegreen 2000a). This is
unlike the red globular cluster populations, which correlate well with
the host galaxies. Milky Way globular clusters are even bluer than the
blue globulars in ellipticals, and may have formed before the elliptical
globulars or in more metal-free regions. Perhaps these globulars formed
in low mass dwarf-like galaxies at $z>4$, and then came together when
our halo potential appeared (e.g., Searle \& Zinn 1978).

After the globular clusters and spheroid stars collected into the Milky
Way, 90\% of the gas that made them was left over, as inferred from
their low metal abundances.  This and other gas presumably cooled and
condensed, forming the bulge (Zinn 1990; Carney, Latham, \& Laird 1990;
Wyse \& Gilmore 1992).  Such dissipational collapse includes the possible
formation of new globular clusters (Zinn 1990), bar/disk self-regulation
(van den Bosch 1998), the formation of giant disk clumps (Noguchi 1999),
continued satellite accretion (Aguerri, Balcells \&  Peletier 2001; Hammer
et al. 2001), super starbursts (Koppen \& Arimoto 1990; Elmegreen 1999),
superwinds (Efstathiou 2000), and multiphase structure (Molla, Ferrini,
Gozzi 2000).  A comprehensive model for the Milky Way was in Samland,
Hensler, \& Theis (1997).

There is good evidence for dissipational collapse in our Galaxy bulge.
In a survey of bulge K giant stars, Minniti (1996) found that as the
metallicity increased for presumably younger stars, the random speed
decreased and the rotation speed increased.  This directly shows the time
sequence of collapse and spin-up for successive generations of stars
during bulge formation.  In addition, Aguerri, Balcells \&  Peletier
(2001) found that bulge growth by accretion changes an initial exponential
light profile to an $r^{-1/4}$ profile, yet the Milky Way bulge profile is
best fit by an exponential (Kent et al. 1991).  Thus the Milky Way bulge
probably did not form by accretion or coalescence of smaller spheroids,
but by dissipational collapse from the halo.  Other galaxy bulges may
have formed by accretion (Hammer et al. 2001), so there could be several
formation mechanisms for bulges.

Following the formation of the bulge, the Milky Way disk gas presumably
built up from continued infall and minor-merger accretions (at $z<2$).
Stars formed in the disk by gravitational processes like what we see
today, involving a threshold column density, dynamical timescales,
a predominance of clusters, and a``Universal'' IMF (e.g., Fall \&
Efstathiou 1980).

\section{Closed Box versus Hierarchical Models}

Closed box or monolithic collapse models (e.g., Eggen, Lynden-Bell \&
Sandage 1962) have often been used to model the evolution of metallicity
and galaxy luminosity functions. Usually there is an epoch of high
mass accretion, possibly until $z\sim2$ or 3, and then a quiescent,
semi-isolated period which is the``closed box.''  Recent models like
this are in Guiderdoni et al. (1998), Chiappini et al. (1999), Pei
et al. (1999), Somerville \& Primack (1999), Bell \& Bower (2000),
Bell \& de Jong (2000), Buonomo et al. (2000), Boissier \& Prantzos
(2000), van den Bosch (2000), Springel (2000), Devriendt \& Guiderdoni
(2000), Prantzos \& Boissier (2000), Boselli et al. (2001), Buchalter
et al. (2001), Boissier et al. (2001), and Rowan-Robinson (2001).

The line fitting the data in Figure 1 is from a closed box model (Boselli
et al 2001).  As part of this model, Boselli et al. also explain why
the ratio of the present star formation rate to the past average rate
correlates with the galaxy mass, as does the current gas mass fraction.
The primary assumption is that the star formation time is shorter in
more massive galaxies (see also Boissier \& Prantzos 2000).

The observation of hyperluminous infrared galaxies requires early star
formation and suggests approximate closed box evolution in subsequent
times. Hyperluminous galaxies have star formation rates of $\sim1000$
M$_\odot$ yr$^{-1}$. Rowan-Robinson (2000) and Pearson (2001) have
identified ultra-luminous infrared galaxies at high z with
$L\sim10^{12}$ - $10^{13}$ L$\odot$. If a galaxy forms stars on a
dynamical time, which is about as fast as it can, then for
$\rho\sim1$ cm$^{-3}$, which is the tidal density limit, and velocity
dispersion $c\sim250$ km s$^{-1}$ from the potential, the mass of $M\sim
\rho\times$Vol$\sim10^{11}$ M$_\odot$ forms in about a crossing time,
$\left(G\rho\right)^{-1/2}$, and this gives a star formation rate of
$\sim c^3/G\sim1000$ M$_\odot$ yr$^{-1}$. Triggers for such rapid and
global star formation could be interactions between giant gas clouds
that leave elliptical galaxies after the stars mix (see review in
Sanders \& Mirabel 1996).

Observations of galaxies with such high star formation rates imply that
the mass was assembled quickly, not slowly from pieces that already
formed their stars at lower rates. Ben\'itez et al. (1999) claim, for
example, that ellipticals formed at $z>5-10$. We have not seen these
proposed young galaxies yet, but extinction corrections implied by
infrared observations could be large enough to hide them (e.g. Blain
et al. 1999; Ramirez-Ruiz et al. 2000). If there is a significant
population of massive bright galaxies at high $z$, forming the
ellipticals and S0's, for example, then these galaxies would presumably
evolve thereafter according to the closed box models. The fraction of
galaxies that evolve this way, and the fraction of the total time most
galaxies spend in semi-isolation, is not yet known.

Indirect evidence for early galaxy formation is that elliptical galaxies
correlate with QSOs.   The redshift evolution of QSOs and bright AGN's is
the same as for elliptical galaxies, and the evolution of weak AGNs is
the same as for spiral galaxies (Boyle \& Terlevich 1998; Franceschini
et al. 1999; Cen 2000).  Thus, ellipticals formed quickly and early,
perhaps by rapid interactions, and not slowly by hierarchical building
over several billion years.

A further test of the closed box model is disk fading with decreasing z.
From $z=1.2$ to $z=0.5$, disks fade by 1.2 mag in blue passbands (Schade
et al. 1995; Brinchmann et al. 1998). If disks fade at $z<1$, then they
are not growing much by hierarchical accretion. Hierarchical buildup
requires evolution of the galaxy number density, but at $z<1$, such
evolution is observed primarily for low mass or blue galaxies
(Glazebrook et al. 1995; Lilly et al. 1995; Ellis et al. 1996;
Brinchmann et al. 1998). In general, peculiar galaxies show more
evolution at $z<1$ than spirals or ellipticals (Brinchmann \& Ellis
2000). Steidel et al. (1999) found no change in the general galaxy
luminosity function out to z=4, and Shanks et al. (2001) found no change
in the number density of bright galaxies out to $z=6$. There is
apparently no change in the density of elliptical galaxies either, out
to at least $z\sim2$ (Ben\'itez et al. 1999).

These observations support the case for early galaxy formation followed by
nearly closed-box evolution.  They contrast with some predictions of big
bang models in a $\Lambda$CDM cosmology.  These models form small dark
matter clumps early on, and then rely on prolonged coalescence to get
L$^*$ galaxies.  The hierarchical models are extremely successful in some
respects: they get the galaxy luminosity function, the DLA distribution,
and the observed level of galaxy interactions at high z.  However, the
galaxy and cluster core densities are often too large in these models
(which need SF feedback or dynamical heating), and the model disks are
too small.  Recent studies are in Mo, Mao \& White (1998), Weil, Eke, \&
Efstathiou (1998), Col\'in et al. (1999), Navarro \& Steinmetz (2000),
Firmani \& Avila-Reese (2000), Koda, Sofue, \& Wada (2000), and Hatton \&
Ninin (2001).

Observational evidence for hierarchical build-up comes from the
metallicity, which is too low at high z to have had most star formation
take place at a very early time (Madau et al. 1998).  If most galaxies
formed early and had a decaying star formation rate thereafter, then
the metal abundance in DLA absorbers should be much higher than observed
(Pettini et al. 1997). A hierarchical model with smaller star formation
rates at high z fits the metallicity observations better.  We cannot
be sure, however, that the DLA metallicities represent the average in
the Universe at their redshifts.  Elliptical galaxies could have formed
early and produced a lot of metals that did not get into the DLA systems.

The K-band luminosity function at high z (Songaila et al. 1994; Cowie
et al. 1996) also agrees with the hierarchical model better than a model
with pure luminosity evolution (Kauffmann \& Charlot 1998).  This implies
there are too few galaxies at high z for pure luminosity evolution.
There is some uncertainty about extinction at high z, though. Extinction
can hide galaxies and decrease the luminosity function.

Other evidence for hierarchical build-up is that the brightest cluster
galaxies, which are mostly ellipticals, get bluer but not brighter with
increasing z. This requires half the mass at z=1 for these galaxies
compared with their masses today (Aragon-Salamanca et al. 1998; McCracken
et al. 2000).  Alternatively, there could be a low-mass bias to the
stellar IMF (Broadhurst \& Bouwens 2000; McCracken et al. 2000).

In summary, the Milky Way roughly fits into the cosmological formation
scenario observed at high z: the spheroid and globular clusters formed
at $z>3$, the bulge at $z\sim2-3$, and the disk at $z<2$. The closed box
and hierarchical models are debated because of unknown extinction
corrections at high z, possible IMF variations, high z selection
effects, and field-to-field variations. Apparently, some galaxies,
particularly the ellipticals and maybe also some giant spirals, formed
quickly and early, perhaps in a hierarchical fashion, and then decayed
in semi-isolation as in the closed box models. The Milky Way bulge and
disk may be examples of closed box evolution, perhaps because of the low
density of other galaxies in our neighborhood. Smaller galaxies
continued to coalesce or merge with larger galaxies up to modern times.
Implicit in this closed-box scenario is the requirement that the
metallicity distribution in the early Universe was not uniform: metals
forming early in elliptical galaxies could not have dispersed very far.
The DLA absorbing clouds then have to be more primordial than the
average cosmic matter.

\section{What can we Learn from Local Star Formation?}

The 2MASS survey suggests that 90\% of local star formation is in dense
clusters (e.g., Lada et al. 1991; Carpenter 2000), having  $\sim10^4$
stars pc$^{-3}$.  Approximately 0.5\% of the gas participates in this
cluster formation inside a molecular cloud (e.g., McLaughlin 1999).
The gas structure is hierarchical and fractal prior to collapse,
presumably as a result of turbulence and self-gravity (e.g., Falgarone \&
Phillips 1990), and the stars form in fractal patterns (Motte, Andre \&
Neri 1998; Testi et al. 2000; Elmegreen \& Elmegreen 2001) in only a few
crossing times (Ballesteros-Paredes, Hartmann, \& Vazquez-Semadeni 1999;
Elmegreen 2000b).

Hierarchical and fractal structure can explain the cluster mass function,
$ n(M)dM\propto M^{-2}dM$ (Fleck 1996; Elmegreen \& Falgarone 1996;
Elmegreen \& Efremov 1997).  This mass function is appropriate for
galactic clusters (Battinelli et al. 1994), OB associations (from the
HII region luminosity function: Kennicutt, Edgar, \& Hodge 1989; Comeron
\& Torra 1996; Rozas, Beckman \& Knapen 1996; Feinstein 1997; McKee \&
Williams 1997; Oey \& Clarke 1998), super-star clusters (Whitmore \&
Schweizer 1995; Zhang \& Fall 1999), and high mass halo globular clusters
(Ashman, Conti, \& Zepf 1995).

The structure can give the stellar mass function too if the
clumps turn into stars at the local dynamical rate (Elmegreen
1997; S\'anchez \& Parravano 1999).  This stellar function is
$n(M)dM\propto M^{-2.35}$ dM down to the thermal Jeans mass, which is
$M_J\sim0.3M_\odot\left(T/10\;{\rm K}\right)^2\left(P/10^6\;{\rm km \;
cm}^{-3}\right)^{-1/2}$ for temperature $T$ and pressure $P$.

The difference between a globular cluster and an open cluster is
primarily in the mass. Their average densities are about the same.
Pressure is an important consideration for cluster mass.  A virialized
cluster with $c^2\sim GM/5R$ has a pressure $P\sim0.1 GM^2/R^4$,
giving \begin{equation}M\sim 6\times 10^3 {\rm M}_\odot \left(P/10^8
k_B\right)^{3/2}\left(n/10^5\;{\rm cm}^{-3}\right)^{-2}.\end{equation}
The normalization here is from the Orion core (Lada, Evans, \& Falgarone
1997), considering a stellar density of $10^4$ M$_\odot$ pc$^{-3}$ at
50\% efficiency.  Evidently, in higher pressure environments, the cluster
masses are larger.  This goes a long way toward explaining the primary
differences between halo globular clusters and modern disk clusters:
the pressure in the halo when these globulars formed was very high as
a result of the high gas density and the high velocity dispersion from
the galaxy potential and big bang turbulent motions.

Pressure is also related to the star formation rate.  For a general
disk, $P\sim G\Sigma_{gas}\Sigma_{stars}\propto \Sigma_{gas}^2$.
But the star formation rate scales as  SFR$\propto\Sigma_{gas}^{1.4}$
(Kennicutt 1989, 1998). Thus $P\propto {\rm SFR}^{1.4}$.  Also, from the
previous expression for maximum cluster mass as a function of pressure,
$M_{max}\propto P^{1.5} \propto {\rm SFR}^2$ when clusters all have
about the same stellar density.  The total cluster luminosity is the
integral over $Mn(M)dM$, which, for $\int_{M_{max}}^{\infty}n(M)dM=1$,
equals $M_{max} \ln \left(M_{max}/M_{min}\right)\propto
{\rm SFR}^2$.  Thus the {\it fractional} SF luminosity in the form of
clusters is proportional to the star formation rate, as found by Larsen \&
Richtler (2000).

The high pressure of star formation in the early Universe contributed
to three major differences compared to today's star formation: the
maximum cluster mass increased, forming globular clusters at the high
mass end, the star formation rate increased, perhaps giving the observed
increase in the star formation rate history (Fig. 1), and the fraction
of all star formation in the form of dense massive clusters increased,
giving a prominent population of halo globulars. These changes are
observed in modern starbursts too because of the same high pressures.
There should also be halo field stars along with the old halo globulars,
and intergalactic blue field stars corresponding to the blue globulars
near elliptical galaxies.

Aside from this pressure difference, halo globular cluster formation
looked pretty normal:  the globulars have a normal mass distribution
function for stars (Paresce \& De Marchi, 2000), normal masses for the
expected halo pressure, a normal cluster mass function (at the high
mass end where most of the clusters survived), a normal efficiency of
cluster formation in gas clouds (~0.25\%-0.5\% -- see McLaughlin 1999),
and normal cluster core densities.

There is a feature of modern star formation that has not yet been seen
at high z, and that is a threshold gas column density.  For modern disks,
$\Sigma_{gas}\ge 5$ M$_\odot$ pc$^{-2}$ for a cool phase of interstellar
matter (Elmegreen \& Parravano 1994), and $\Sigma_{gas}\ge0.7\kappa
c/\pi G$ for self-gravity to be important (Toomre 1964; Kennicutt 1989).
Modern disks also have a SFR/Area$\sim0.033\Sigma_{gas}\Omega_{orb}$
(Kennicutt 1998) for orbital angular rate $\Omega_{orbit}$. This areal
rate along with $Q=1$ gives a total rate of star formation equal to
$\left({\rm SFR}/{\rm Area}\right) \times \pi R^2 \sim 0.04 V^2 c / G
\sim 5$ M$_\odot$ yr$^{-1}$ in a typical disk.  This rate corresponds to
a lower efficiency than the value of $c^3/G$ found above for a spheroid
(where $c=V$).

\section{Summary}

Cosmological star formation is observed directly for $z<5$, although
extinction problems, selection effects and survey limitations are
probably important at high $z$.  What is observed to be forming at $z>3$
are primarily spheroidal systems, and these correlate well with the onset
of the QSO phase. The galaxy buildup process is still uncertain, however;
it could have been hierarchical or monolithic, or some of each, depending
on epoch, environment, galaxy mass, and local density.  Galactic bulges
can apparently form by either closed-box or hierarchical scenarios,
although the Milky Way bulge seems to have formed as a closed box. Disks
require quiet environments in order to survive, and this tends to limit
them to $z<2$.  The source of the disk gas is not known.

Globular clusters look normal in comparison to observations of present
day star formation. The primary difference seems to have been the high
pressure of their environment, which is expected for galactic halo
conditions. These high pressures increased the maximum cluster mass,
the star formation rate, and perhaps the cluster formation efficiency.

\end{document}